\documentclass[12pt]{article}
\usepackage{amsmath}
\usepackage{graphicx}
\usepackage{tabu,multicol,multirow,booktabs}
\usepackage{enumerate}
\usepackage{natbib}
\usepackage{url} 

\begin{document}

\title{Bayes-ically fair: A Bayesian Ranking of the Olympic Medal Table}
\author{
  C. MacDermott, C.J. Scarrott, J. Ferguson \\
  School of Mathematical and Statistical Sciences, University of Galway
}
\maketitle

\begin{abstract}
Evaluating a country’s sporting success provides insight into its decision-making and infrastructure for developing athletic talent.  The Olympic Games serve as a global benchmark, yet conventional medal rankings can be unduly influenced by population size. We propose a Bayesian ranking scheme to rank the performance of National Olympic Committees by their `long-run' medals to population ratio. The algorithm aims to mitigate the influence of large populations and reduce the stochastic fluctuations for smaller nations by applying shrinkage. These long-run rankings provide a more stable and interpretable ordering of national sporting performance across games compared to existing methods.
\end{abstract}


\section{Introduction}
\label{sec:introduction}

The Olympic Games represent the pinnacle of international sporting competition, offering a global stage to evaluate national athletic success. Every four years, elite athletes from nations and territories represented by National Olympic Committees (NOCs) compete across a wide variety of sports for `Olympic glory'. As such, medal tables, comparing the relative performance of differing NOCs, are examined by media outlets and sports authorities world-wide. Particular scrutiny goes on how many medals their home country has won, where they rank compared to other nations, and how their performance compares with previous Games. 

During the 2024 Olympic Games in Paris, one of the key narratives centered on whether the United States or China would top the standard lexicographic ordering of the medal table, which ranks nations primarily by gold medal count. Both finished with 40 gold medals, but the United States was placed ahead due to its higher overall medal tally.

However, while total medal counts are a common point of media focus, they inherently favor more populous NOCs, as these have a larger pool of athletes to draw from. For example, Ireland will never top a medal table based purely on total counts, as the sheer disparity in population makes it virtually impossible to compete with nations hundreds of times larger. To address this imbalance, an alternative approach is to scale medal counts by population, producing the so-called `per-capita' table.  However, these tables are also crude in that they fail to recognize inherent stochasticity over repeated Olympic cycles, which on a relative scale will be large for smaller countries.  The result is that the per-capita table is typically led by small population countries.  For instance, Grenada, Dominica and Saint Lucia, with respective populations of 112,579; 67,408 and 184,100 occupied the top 3 positions on the 2024 per-capita table.  

In response to these limitations, approaches have been proposed in the scientific literature to create alternative rankings. These approaches can be divided into methods that challenge the standard ranking produced by the number of gold medals (or total medals) by (1) reweighing the contributions of gold, silver and bronze medals in various ways \citep{jrnl:ogwang2021olympic}, (2) producing composite tables combining medal tables ordered by differing criteria \citep{jrnl:cao2016measuring}, and (3) ranking performance relative to collective resources available to a particular country, as specified by variables such as population and GDP using approaches like Data Envelope Analysis \citep{jrnl:lozano2002measuring,jrnl:bernard2004wins,jrnl:churilov2006towards,jrnl:wu2009dea,jrnl:li2015performance,jrnl:flegl2018measuring}, and linear regression \citep{jrnl:halsey2009true}.  

Finally, the method recently proposed by \cite{jrnl:duncan2024population} has received substantial attention in various popular media outlets around the world, like the The New York Times \citep{news:NYTDuncan} and The Guardian \citep{news:GuardianDuncan} who referred to it as the Goldilocks method due to not favoring large or small countries. The Duncan-Parece method ranks according to a novel probabilistic index that adjusts for population effects, which we will compare and contrast to our own approach below.

In this article, we propose Bayesian ranking as an alternative method to order medal tables. Unlike the traditional approaches, Bayesian ranking focuses on the most likely ordering of the key parameters of interest that drive the observed data \citep{jrnl:laird1989empirical}.  In the Olympic medal setting, we can informally define these key parameters as the long-run per-capita medal ratios that would be observed in differing countries if the Olympics were repeated several times under differing conditions.  In other words, the medal table we produce via Bayesian ranking does not aim to summarize what happened in a particular Olympic Games – but instead to infer which countries are most efficient in producing Olympic medalists, relative to their population size over the long term.  

In the case that the data from a single Olympics is used, the ranking produced could be considered a modification of the ranking by the observed per-capita medal ratios. For large countries such as the USA and China, the Bayesian estimates of long-term per-capita ratios will closely resemble observed values, given their large populations and relatively stable performance. In contrast, for smaller countries, the Bayesian estimates may diverge significantly from observed ratios, as their medal counts are more susceptible to stochastic variation within a single Olympic cycle.  

A key consideration in this approach is the choice of prior distribution for these underlying long-term ratios, as the reliability of the rankings relies on how reasonable the prior is. Hierarchical Bayesian modeling, as used in the methods discussed later, mitigates some subjectivity by leveraging the observed distribution of medal counts across countries to help estimate the prior, thereby enhancing the robustness of the resulting rankings. 

In Section~\ref{sec:methods}, we will describe our Bayesian ranking model in detail, with careful discussion regarding the plausible validity of our modeling assumptions.  In Section~\ref{sec:results}, we examine the medal tables produced by our approach over the last six Olympic cycles, and contrast with the results produced by the recent algorithm proposed by \cite{jrnl:duncan2024population} mentioned above. In Section~\ref{sec:discussion}, we provide additional evaluation and justification of our modeling assumptions, and discuss both caveats as well as possible extensions of our approach.

\section{Methods}
\label{sec:methods}

The National Olympic Committees (NOCs) represent each country, as well as the individual athletes from the Refugee Olympic Team. The $C$ NOCs that competed at a games are indexed by $c=1, \ldots, C$ and the total population size per NOC is given by $n_c$. These vary in each Olympic Games, but for notational convenience we'll ignore this in the following presentation which will focus on the NOCs in Paris 2024  ($C=204$). As with the usual per-capita analysis, only NOCs with a measurable population are included.  The total number of medals won by country $c$ is the sum of the number of medals won by their individual athletes $M_c=1 \times M_{1,c}+2\times M_{2,c}+\cdots$, where $M_{i,c}$ is the random variable for the number of their athletes that won exactly $i$ medals. Here each medal-winning country in team competitions such as relays or field sports is considered a nominal athlete that wins a single medal, which is in keeping with the official lexiographic ranking compiled by the International Olympic Committee. The total number of medals achieved by each country $c$ will be modeled using a combination of interrelated Poisson processes representing the number of athletes $M_{i,c}$ winning exactly $i$ medals.

If we momentarily ignore the number of total medals, $M_c$, and consider only the number of unique medal winning athletes, $U_c=\sum_i M_{i,c}$, we can assume $U_c$ to be roughly Poisson distributed. The Poisson distribution is used as an approximation to the conceptual idea that the number of unique medal winners from country $c$ is approximately a $U_c\sim\textrm{Binomial}(n_c, p_c)$ distribution, where a unique medal winner is considered a rare event from a large number $n_c$ of independent Bernoulli trials representing each person within a population with probability of success $p_c.$ The binomial/Poisson assumptions are considered in Section~\ref{sec:discussion}. 

However, care has to be taken with modeling total medals won, $M_c$, as the world's highest achieving athletes that win more than one medal would invalidate the Poisson limiting assumption that it should be impossible to have two or more events at the same instance. Hence, we will instead model the number of athletes winning $i=1, 2, \ldots$ as conditionally independent Poisson processes, that are interlinked via a prescribed relationship between the country-specific impacts on the probability of winning that number of medals. Note that this assumption implies that the number of unique medal winners, $U_c=\sum_i M_{i,c}$ \textit{will} follow a Poisson distribution, whereas the total number of medals won $M_c=1 \times M_{1,c}+2\times M_{2,c}+\cdots$, will not.

In Paris 2024 the highest number of medals won by any individual athlete was four medals, so for notational convenience we will denote $M_{4,c}$ as the number of athletes winning four or more medals, and so we only consider the finite sum to $M_{4,c}$. It is straightforward to adapt the modeling framework to allow for a different number of medals won by an individual athlete. The number of athletes winning $i=1, 2, 3$ and 4 or more medals in each competing country are assumed to be Poisson distributed:
\begin{eqnarray*}
  M_{1,c} &\sim& \textrm{Poisson}(\lambda_{1,c}),\\
  M_{2,c} &\sim& \textrm{Poisson}(\lambda_{2,c}),\\
  M_{3,c} &\sim& \textrm{Poisson}(\lambda_{3,c}), \textrm{and}\\
  M_{4,c} &\sim& \textrm{Poisson}(\lambda_{4,c}).\\
\end{eqnarray*}
The expected number (and variance) of athletes winning each number of medals is then $\lambda_{i,c}=n_c\,p_{i,c}$, where $p_{i,c}=P(X_c=i)$ is the probability of an individual $X_c$ from competing country $c$ winning exactly $i$ medals. Again, for convenience in notation $p_{4,c} = P(X_c\geq 4)$.  The country-specific Poisson distributions are equivalent to a classical frequentist inference of a country-specific random effect on the probability of a winning that number of medals.

These Poisson distributions are treated as independent of each other, conditional on those country-specific probabilities $p_{1,c}, p_{2,c}, p_{3,c}$ and $p_{4,c}$ of an individual winning a given number of medals. The relationship between these probabilities is defined through parameters for the following conditional probabilities:
\begin{eqnarray*}
    p_{c}&=&P(X_c\geq 1) \text{ i.e., the probability of athlete from $c$ being a unique medal winner},\\
    q_2&=&P\left(X\geq 2\,\vert\,X\geq 1\right),\\
    q_3&=&P\left(X\geq 3\,\vert\,X\geq 2\right), \textrm{and}\\
    q_4&=&P\left(X\geq 4\,\vert\,X\geq 3\right).
\end{eqnarray*}
The probability of being a unique medal winner $p_c$ retains the country-specific subscript notation. In contrast, the consecutive conditional probabilities $q_i$ of winning at least one additional  medal having already won $i$ medals are not country-specific, once the conditioning on being a unique medal winner from that NOC is taken into account. There is a slight abuse of notation here in that these conditional probabilities depend on the country-specific random variable $X_c\geq 1$ for being unique medal winner, but we use the random variable $X$ to represent the number of medals won by an individual from a generic NOC winning that number of medals.

The country-specific effect of population biology, demography, and the ability to develop sporting talent is assumed to be entirely driven through the parameter for the chance of producing a unique medal winner $p_c=P(X_c\geq 1).$ The above framework then assumes that conditional on someone being a medal winner, the likelihood of winning further medals has no further country-specific effects. This assumption seems reasonable as when conditioning on the rare event of winning one or more medals, we are implicitly conditioning on an athlete that has the talent, training environment and funding to perform at a world class level in their chosen sport; the conditional probability of winning more than one medal given this context might be largely independent of the athlete’s home country. These multi-medal winning athletes are also sufficiently rare that they will not be very influential on the rankings inference, except for the small countries that have been fortunate enough to host such talent. Indeed we see no substantive change in the results presented when only the unique medal winning athletes are modelled.

From a computational view this construction has the benefit of not needing country specific random effects for these multi-medal winning probabilities which would be hard to inform from the data as the majority of countries have no such athletes. The country-specific probabilities of being a unique medal winner are more reliably estimated as there are based on a larger number of such athletes.

The unconditional probability of an individual from country $c$ of winning a certain number of medals will vary by country according to:
\begin{eqnarray*}
    p_{1,c} = P(X_c=1)&=&P(X_c\geq 1)- P(X_c\geq 2)=p_c\,(1- q_2),\\
    p_{2,c} = P(X_c=2)&=&P(X_c\geq 2)- P(X_c\geq 3)=p_c\,q_2\,(1-q_3),\\
    p_{3,c} = P(X_c=3)&=&P(X_c\geq 3)- P(X_c\geq 4)=p_c\,q_2\,q_3\,(1-q_4), \textrm{and}\\
    p_{4,c} = P(X_c\geq 4)&=& p_c\,q_2\,q_3\,q_4.
\end{eqnarray*}
These parameters are given uninformative and independent prior distributions of
\begin{eqnarray*}
    q_2&\sim& \textrm{Uniform(0,1)},\\
    q_3&\sim& \textrm{Uniform(0,1)}, \textrm{and}\\
    q_4&\sim& \textrm{Uniform(0,1)},
\end{eqnarray*}
so that they are predominantly learnt from the data. The country-specific probability of an individual becoming a unique medal winner is assumed to follow a prior distribution:
$$p_c\sim \textrm{Beta}(\alpha, \beta),$$
where $\alpha$ and $\beta$ are given independent uniform hyperprior distributions:
\begin{eqnarray*}
    \alpha&\sim& \textrm{Uniform}(0, 1), \textrm{and}\\
    \beta&\sim& \textrm{Uniform}(0, 10^8).
\end{eqnarray*}

Note that sensible values for $\alpha$ and $\beta$ should correspond to an expected medal probability from the associated beta-prior: $E(p_c)=\frac{\alpha}{\alpha+\beta}$, that is relatively close to the total number of medals divided by the global population ($1039/8,000,000,000=1.3\times10^{-7}$ for Paris 2024).  However, this information has not been hardcoded into the model: instead, the non-informative hyperpriors give the model the flexibility to learn the most plausible values for $\alpha$ and $\beta$ based on the observed medal counts across differing countries.

The priors induce shrinkage to reduce the uncertainty for small countries which have limited sample information by pulling their estimates towards the global average, while preserving the data-driven estimates for larger countries. The pooling of information across all countries for the multi-medal winning athletes also ensures they do not have undue influences on the estimates for small countries. The shrinkage provides a better estimate of the `long-run' probability of winning medals, which would provide a more stable ranking of each nation's performance. Alternative prior specifications are considered in the online Appendix, with little sensitivity to this choice found.

In summary, we have just four parameters to obtain a posterior distribution for each country; namely the country-specific random effect for the probability of a unique medal winner $p_c =P(X_c\geq 1),$ and the three conditional probabilities of such an athlete winning further medals $q_i = P\left(X\geq i\,\vert\, X\geq i-1\right)$ for $i=2, 3$ and $4$.

The population size data are obtained from the United Nations \cite{web:UN_Data} which are updated for each Olympic year. The Olympics data are obtained from the \cite{web:IOC_data}. Gibbs sampling using the \texttt{rJAGS} package \citep{man:rjags} is used to draw posterior samples of $(p_c, q_2, q_3, q_4)$ for each country, which are then transformed as above to country-specific medal probabilities, and subsequently to expected per-capita medal ratios: $E(M_c/n_c)=p_{1,c} + 2\,p_{2,c} + \cdots$. The rank for the expected per-capita number of medals won in a particular country, among all medal-winning countries, is then calculated for each posterior sample.  Finally, the estimated posterior mean of these ranks is used to order the medal table as is commonplace in the Bayesian ranking literature. Alternative statistics, like the posterior median rank, that could be used for the ranking are considered in Section~\ref{sec:discussion}.

Finally, it is worth noting that while we base our ranking on the posterior distribution of the expected medal rate per capita $E(M_c/n_c)$ in the above, this is equivalent to ranking on the probability of an individual from that country winning at least one medal $p_c=P(X_c\geq 1),$ which is easily seen by expansion of expectation:
\begin{eqnarray*}
  E(M_c/n_c)&=&E(M_{1,c}/n_c) +2\,E(M_{2,c}/n_c)+3\,E(M_{3,c}/n_c)+4\,E(M_{4,c}/n_c)\\
            &=& p_c\,(1-q_2 + 2\,q_2\,(1-q_3) + 3\,q_2\,q_3\,(1-q_4) + 4\,q_2\,q_3\,q_4)
\end{eqnarray*}
where the terms in the bracket are not country specific.

\section{Results}
\label{sec:results}

\begin{table}
\begingroup\fontsize{8}{10}\selectfont
\resizebox{\linewidth}{!}{%
\begin{tabu} to \linewidth {>{\raggedleft}X[0.8] >{\raggedleft}X[0.7] >{\raggedleft}X[1.1] >{\raggedleft}X[0.3] >{\raggedright}X[1.9] >{\raggedright}X[0.3] >{\raggedleft}X[1.6] >{\raggedright}X[0.9] >{\raggedright}X[0.9] >{\raggedright}X[1.2] >{\raggedright}X[0.8] >{\raggedright}X[1]}
\toprule
\multicolumn{3}{c}{\textbf{Ranks}} & \multicolumn{5}{c}{\textbf{Country}} & \multicolumn{2}{c}{\textbf{Medals per million}} & \multicolumn{2}{c}{\textbf{Credible Interval}}\\
\cmidrule(lr){1-3} \cmidrule(lr){4-8} \cmidrule(lr){9-10} \cmidrule(lr){11-12}

Posterior Mean & Per-capita & $U$-Index & & Name & Code & Population (thousands) & Medal Total & Observed & Posterior Median & 95\% Lower & 95\% Upper\\

\midrule
1 & 4   & 6   & \includegraphics[height=1em, width = 1.5em]{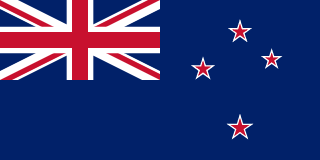}\quad  & New Zealand   & NZL   & 5,214     & 20 & 3.84 & 3.31 & 2.05 & 4.89\\
2 & 7   & 1   & \includegraphics[height=1em, width = 1.5em]{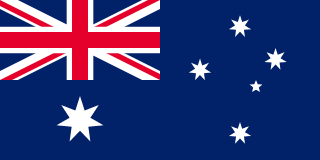}\quad  & Australia     & AUS   & 26,713    & 53 & 1.98 & 1.80 & 1.32 & 2.31\\
3 & 8   & 8   & \includegraphics[height=1em, width = 1.5em]{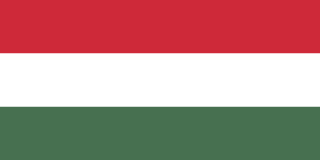}\quad  & Hungary       & HUN   & 9,676     & 19 & 1.96 & 1.83 & 1.06 & 2.89\\
4 & 10  & 4   & \includegraphics[height=1em, width = 1.5em]{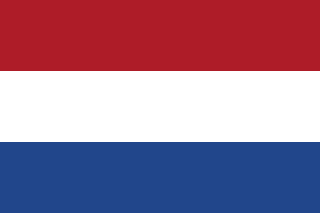}\quad  & Netherlands   & NLD   & 18,229    & 34 & 1.87 & 1.68 & 1.15 & 2.40\\
5 & 1   & 27  & \includegraphics[height=1em, width = 1.5em]{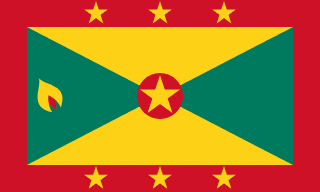}\quad  & Grenada       & GRD   & 117       & 2  & 17.09 & 2.42 & 0.34 & 8.16\\
6 & 6   & 18  & \includegraphics[height=1em, width = 1.5em]{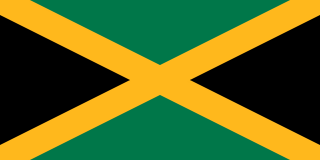}\quad  & Jamaica       & JAM   & 2,839     & 6 & 2.11 & 1.71 & 0.67 & 3.54\\
7 & 5   & 25  & \includegraphics[height=1em, width = 1.5em]{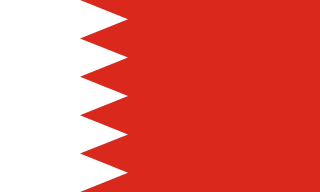}\quad  & Bahrain       & BHR   & 1,607     & 4 & 2.49 & 1.74 & 0.53 & 4.25\\
8 & 11  & 15  & \includegraphics[height=1em, width = 1.5em]{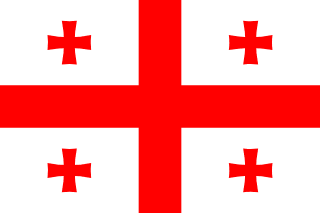}\quad  & Georgia       & GEO   & 3,808     & 7 & 1.84 & 1.58 & 0.65 & 3.00\\
9 & 12  & 17  & \includegraphics[height=1em, width = 1.5em]{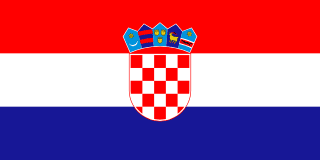}\quad  & Croatia       & HRV   & 3,875     & 7 & 1.81 & 1.60 & 0.65 & 3.10\\
10 & 13 & 13  & \includegraphics[height=1em, width = 1.5em]{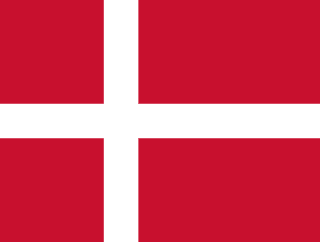}\quad  & Denmark       & DNK   & 5,977     & 9 & 1.51 & 1.35 & 0.64 & 2.50\\
15 & 18 & 20  & \includegraphics[height=1em, width = 1.5em]{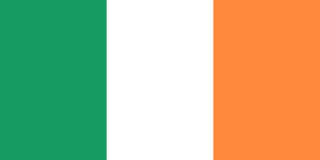}\quad  & Ireland       & IRL   & 5,255     & 7   & 1.33  & 1.01  & 0.42  & 2.04\\
16 & 24 & 2   & \includegraphics[height=1em, width = 1.5em]{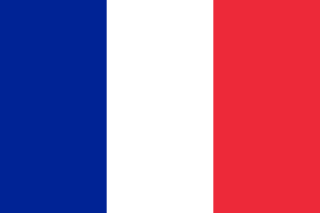}\quad  & France        & FRA   & 66,549    & 64  & 0.96  & 0.95  & 0.72  & 1.22\\
17 & 25 & 3   & \includegraphics[height=1em, width = 1.5em]{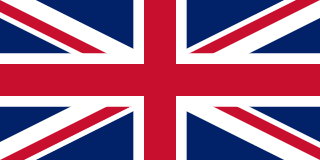}\quad  & Great Britain & GBR   & 69,138    & 65  & 0.94  & 0.95  & 0.72  & 1.21\\
21 & 2  & 41  & \includegraphics[height=1em, width = 1.5em]{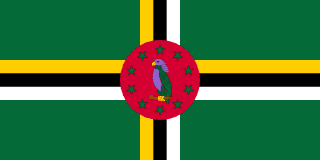}\quad  & Dominica      & DMA   & 66        & 1   & 15.15 & 1.10 & 0.06 & 5.64\\
24 & 3  & 29  & \includegraphics[height=1em, width = 1.5em]{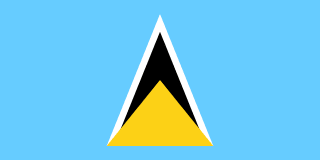}\quad  & Saint Lucia   & LCA   & 180       & 2   & 11.11 & 0.96 & 0.05 & 4.98\\
46 & 42 & 43  & \includegraphics[height=1em, width = 1.5em]{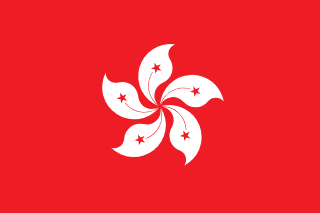}\quad  & Hong Kong     & HKG   & 7,415     & 4   & 0.54 & 0.36 & 0.09 & 0.96\\
51 & 48 & 5   & \includegraphics[height=1em, width = 1.5em]{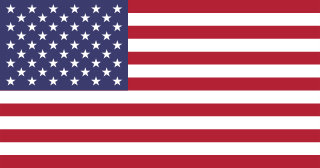}\quad  & United States & USA   & 345,427   & 126 & 0.36  & 0.34  & 0.29  & 0.41\\
76 & 76 & 114 & \includegraphics[height=1em, width = 1.5em]{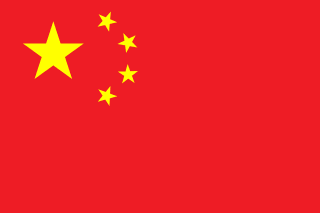}\quad  & China         & CHN   & 1,419,321 & 91  & 0.06  & 0.06  & 0.05  & 0.08\\
\bottomrule
\end{tabu}%
}
\endgroup

\caption{Bayesian ranking of the top 10 and selected countries in Paris 2024 by posterior estimates of medal-winning probabilities, normalized per million population, including 95\% credible intervals. The Duncan-Parece $U$-index based ranks are provided.}
\label{tab:ranks}
\end{table}

Table~\ref{tab:ranks} presents the Bayesian ranking of the top ten and selected countries from Paris 2024. The posterior mean rank over those countries that won at least one medal is used to order the table. As per usual approaches, the NOCs that did not win any medals are not ranked. Population size, total medals won, observed per-capita medal winning rates and rankings by these observed rates, as well as the \cite{jrnl:duncan2024population} $U$-index are also provided for comparison. The posterior median medal winning rate per million and 95\% credible intervals for uncertainty estimation are provided. 

The posterior median is utilized due to the strong skewness in the credible intervals for smaller countries to be discussed in Section~\ref{sec:results-uncertainty}. It is worth noting that the posterior median rate does not follow exactly the same ordering as the posterior ranks due their skewness. A Shiny app is available from \cite{web:shiny}, which allows users to explore the different rankings (including the posterior median rank, as an alternative) across all NOCs, including demographic comparisons for the last six Olympic Games going back to Athens 2004.

\subsection{Bayesian Shrinkage and Ranking Stability}
\label{sec:results-shrinkage}
As mentioned, a key feature of the Bayesian ranking approach is its shrinkage effect, which adjusts for small-sample uncertainty. It prevents countries with exceptionally high per-capita medal counts from being overemphasized from a single-event which may not persist over the long-term. The shrinkage can be seen for Grenada in Table~\ref{tab:ranks} having an observed medals per million of 17 (rank 1 per-capita), which decreased to 2.42 (Bayesian rank 5). Similarly, Dominica and Saint Lucia were ranked second and third on a per-capita basis, but have been shrunk to 21 and 24 respectively in the Bayesian ranking.  Using Bayesian ranking, New Zealand, Australia and Hungary are ranked as the top 3, but under the per-capita rate it would be Grenada, Dominica and then Saint Lucia, a combination which is unlikely to be seen again in future Olympics. The Bayesian shrinkage results in a more stable and interpretable ranking compared to naive per-capita methods, which tend to disproportionately favor smaller nations.

\begin{figure}[!ht]\centering
\includegraphics[width=\linewidth]{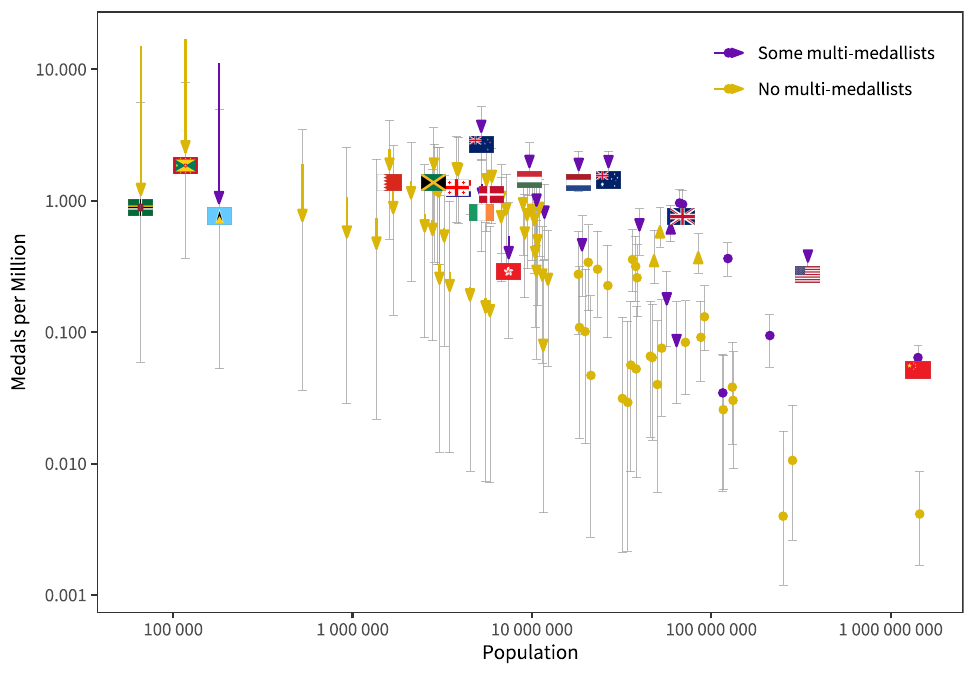}
\caption{Comparison of posterior median and observed medal winning rates by population size for each country with at least one medallist. Directed arrows indicate the shrinkage effect from observed per-capita rates to posterior estimates. The 95\% credible intervals for the medal rate per million are shown by the gray lines, with boundary range ticks. Those countries with any multi-medal athletes are shown as a different color and flags mark countries listed in Table~\ref{tab:ranks}.}
\label{fig:shrinkage}
\end{figure}

The performances of larger countries are well informed by the sample data, so see relatively little change in the posterior inference. The USA and China achieved the same number of gold medals, the USA won both more silver and bronze medals so are ranked first in the official lexicographic medal table. But in the per-capita rankings they are placed as 48 and 76 respectively, due to their very large population sizes. The shrinkage is minimal for such large countries with the the USA moving down 3 positions to 51 in the Bayesian rank and China only 1 rank up to 75. The slightly larger change for the USA is discussed further in Section~\ref{sec:results-multimedal}.

In Figure~\ref{fig:shrinkage} we can see the shrinkage effect across all countries, by comparing the observed and posterior median per-capita medal probabilities. The smaller nations on the left of the graph are pulled closer to the global mean, reflecting the underlying uncertainty in their success rates. In contrast, larger countries on the right see little effect of the prior, as their rankings are more data-driven.

It is worth noting that the level of shrinkage is a function of three factors: (1) the population size, (2) the number of medals won by each NOC (note that these first two factors pertain to the observed medal winning rate) and (3) whether there are any multi-medal winning athletes. This feature can be seen by the shrinkage arrows having different lengths for NOCs with similar population sizes which are close to each other on the $x$-axis. Generally, if two NOCs have same population size and medal counts, the one with more multi-medal winners will see more shrinkage to the global average which will be discussed in Section~\ref{sec:results-multimedal}. The NOCs with an unusually high medal count, for a given population size, will see more shrinkage as is expected when providing more stable medal winning rate estimates. 

It should be noted that only medal winning countries are shown in Figure~\ref{fig:shrinkage}. However, a consequence of the shrinkage is that NOCs that won no medals exhibit shrinkage upwards, which is again to be expected to provide more stable medal winning rate estimates. This shrinkage effect for non-medallist NOCs is discussed in more depth in the online Appendix. It is of course important to include the non-medal winning NOCs in the inference, as they inform the global average rates and their exclusion would lead to a bias.

\subsection{Impact of Multi-Medal Winners on Bayesian Rankings}
\label{sec:results-multimedal}

Another feature of the Bayesian ranking model is its emphasis on the number of unique medal winners (and the number of medals they won) rather than the total medal count. This feature of the model is a notable improvement over \cite{jrnl:duncan2024population} which ignores the athletes that compete in several events and break the binomial assumption that medals won in differing events are independent. The focus on the unique medal winners also helps prevent a small number of extraordinary multi-medalists from disproportionately inflating a country’s ranking. One must not diminish the achievement of such athletes, but a goal here is to provide a more stable estimate of each NOC's medal winning rate which may be more reflective of the NOC's ability to develop sporting talent. The outstanding ability of such athletes is likely more driven by their individual genetic factors, and the particular events they compete in (e.g. more multi-medal winners are seen in swimming events), as opposed to being a strong indicator of an NOC's talent development capability.

This effect is illustrated in Figure~\ref{fig:shrinkage}, where countries with similar populations and nearly identical total medal counts so are near each other on the graph (in both $x$ and $y$ direction) exhibit different levels of shrinkage. The countries colored in purple have multi-medal winners exhibit more shrinkage than those with unique medal winners only which are colored in gold. For instance, in 2024 Ireland and Norway with essentially the same populations won 7 and 8 medals respectively. Ireland’s medals were won by six unique medalists, with one athlete winning multiple medals, while Norway’s medals were distributed across eight different athletes.  As another example, Grenada and Saint Lucia are both small nations with similar populations that won two medals in Paris 2024.  They ranked 1 and 3 on the observed per capita table. However, because Saint Lucia's two medals were won by the same athlete, Julien Alfred, their estimated medal rate is shrunk more than Grenada's, whose two medals were won by separate athletes.  Finally, Hong Kong won four medals in 2024, but two of those were won by the swimmer Siobhán Haughey, which explains the larger shrinkage applied to Hong Kong's observed medal ratio compared to similar size countries.  This feature underscores the model’s tendency to favor broader medal distribution across athletes rather than concentrated success among a few high achieving individuals. This feature can also effect the degree of shrinkage applied to bigger countries, albeit to a lesser extent as the absolute amount of shrinkage is smaller for larger countries. For instance, USA moves by 3 positions in the ranking due to their 10 multi-medal winners winning 23 out of a total 126 medals, whereas China only moves by 1 position as they have only 3 multi-medal winners winning 7 out of there 91 medals. 

\subsection{Ranking Uncertainty and Credible Intervals}
\label{sec:results-uncertainty}

\begin{figure}[!ht]\centering
\includegraphics[width=\linewidth]{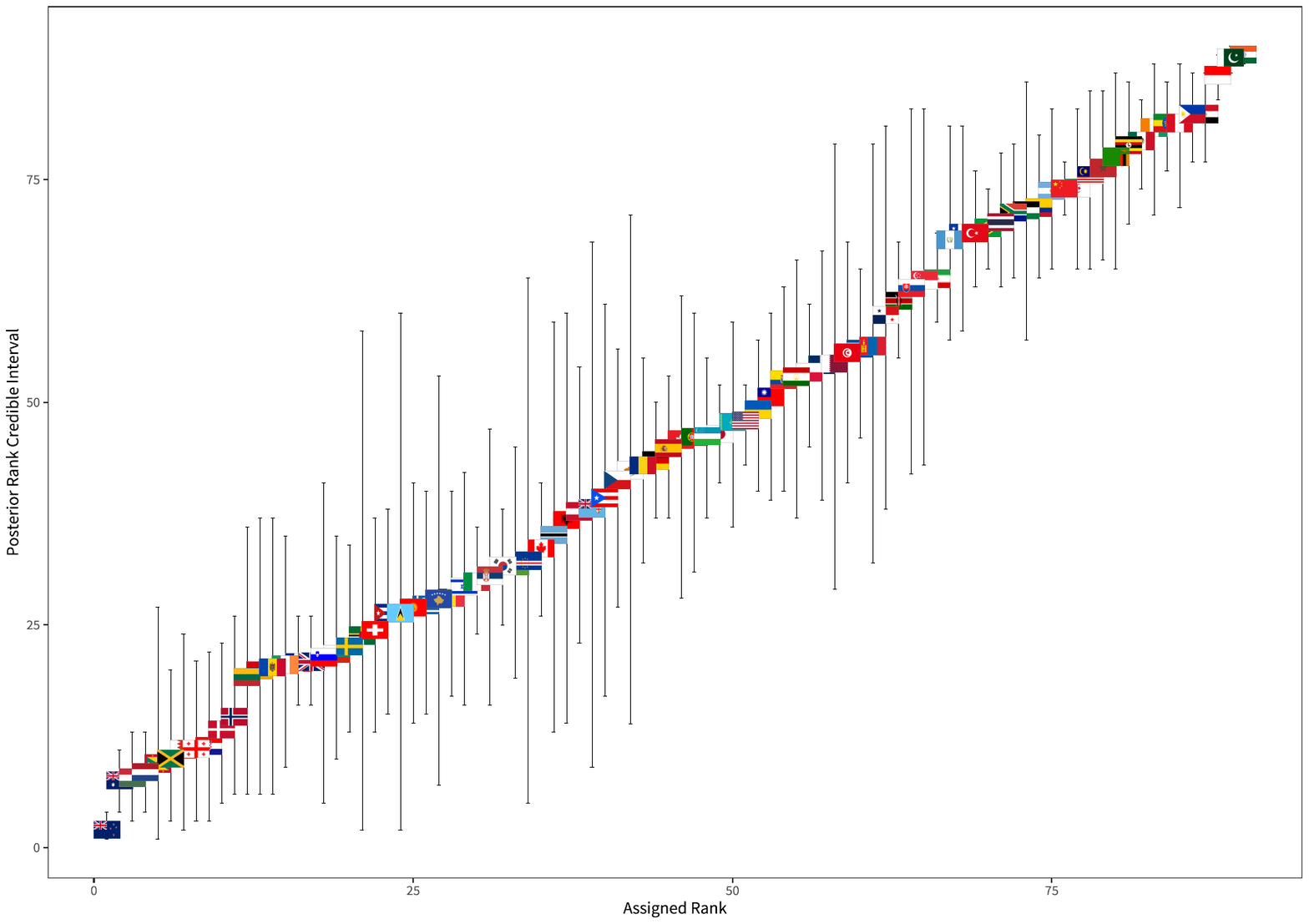}
\caption{Average Bayesian rank with 80\% credible intervals for Paris 2024, ordered by rank. Smaller nations exhibit wider credible intervals due to greater uncertainty in their medal counts, while larger countries have more stable rankings.}
\label{fig:credrank}
\end{figure}

On their own, the posterior mean ranks should not be used to compare the performance of different countries, just like $p$-values should not be used to judge effect sizes, as they do not account for uncertainty. To address this, credible intervals for every medal winning NOC from Paris 2024 in Figure~\ref{fig:credrank}, in addition to the selection in Table~\ref{tab:ranks}, which provide a more nuanced assessment of each NOC’s ranking stability.

As expected, smaller countries exhibit wider credible intervals, reflecting greater ranking uncertainty due to their limited sample sizes, while larger nations have more stable rankings. Further, for well performing countries the rank distributions are right skewed due to the natural bound at first rank and vice-verse for weakest performing countries bounded by the bottom rank. The asymmetry is most clear for the smallest countries with the most uncertainty.

For example, the USA at posterior mean rank 48 has a very short (and symmetric) credible intervals reflecting the large sample of competing athletes. New Zealand’s credible interval does not overlap those of Australia and Hungary, reinforcing its standout position as the top-performing country according to the Bayesian ranking. In contrast, Grenada’s credible interval overlaps substantially with those of many other small nations, suggesting that its superior per-capita medal count may not represent a statistically meaningful performance difference when accounting for sample uncertainty.

In general, countries with fewer medal winners (and smaller populations) exhibit wider credible intervals, while those with consistent performances across multiple events (typically larger nations) have more stable rankings. This highlights a key advantage of the Bayesian approach with the credible intervals - it provides a probabilistic measure of ranking uncertainty, ensuring that countries with overlapping intervals are not misinterpreted as significantly different in performance despite differences in their rank positions.

\subsection{Comparison with Duncan-Parece Method}
\label{sec:results-comparison}

The Bayesian posterior mean ranks based on the estimated medal-winning probabilities presented in Table~\ref{tab:ranks} differ significantly from the \cite{jrnl:duncan2024population} $U$-index rankings. Under the Bayesian ranking, New Zealand, Australia, and Hungary rank among the top-performing nations, while Grenada, despite its high observed per-capita medal count, is ranked  slightly lower due to the shrinkage effects. The USA and China have seen little change in the Bayesian ranking compared to a per-capita rating.

The Duncan-Parece method implicitly penalizes smaller countries in the rankings, as their approach is essentially $p$-value based. It is well known that $p$-values are sample size dependent \citep{jrnl:wasserstein2016asa}. A small sample limits the evidence available, even for a large effect size, so smaller NOCs cannot achieve the small $p$-values required to be ranked highly, unless they attain an extraordinarily large number of medals relative to their population size. As such the larger countries are given an advantage in the Duncan-Parece method.

\begin{figure}[!ht]\centering
\includegraphics[width=\linewidth]{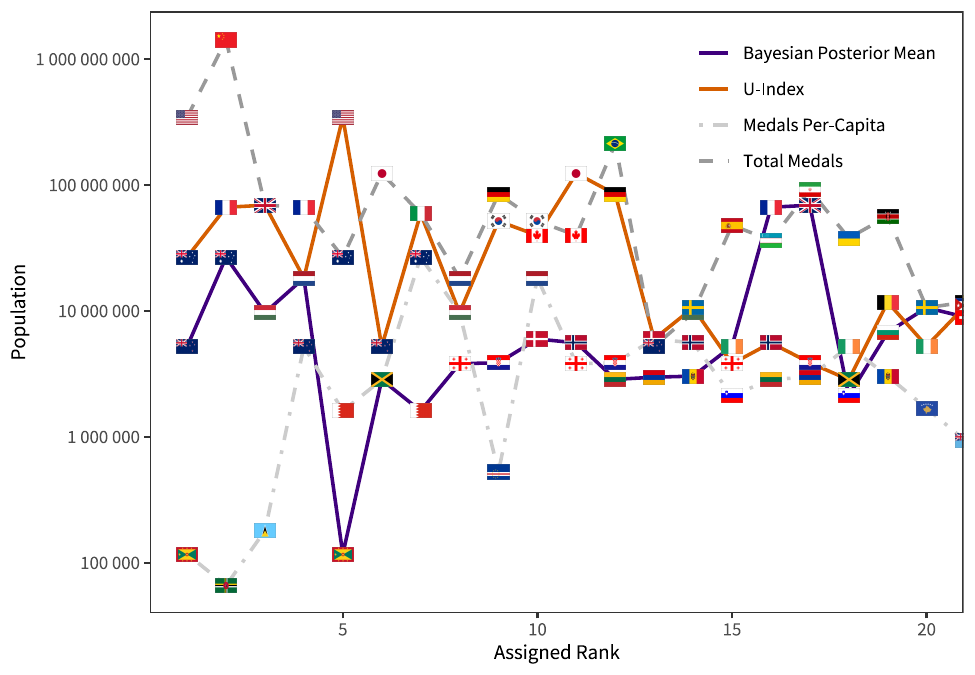}
\caption{Comparison of rankings across different methods for Paris 2024 as a function of population size. Bayesian rank (purple), $U$-index rank (orange), total medals rank (dashed dark gray), and per-capita medals rank (dashed light gray) are shown. Notably, the $U$-index closely follows the total medal count ranking, highlighting its tendency to favor larger nations, while the Bayesian ranking adjusts for population size.}
\label{fig:rankcomp}
\end{figure}

The advantage given to large NOCs in the Duncan-Parece method is seen in Table~\ref{tab:ranks} and Figure \ref{fig:rankcomp} by it assigning higher ranks to large countries like the USA at rank 5, Great Britain at rank 3 and France at rank 2, which benefit from their overall representation of athletes competing across events. In general, the Duncan-Parece $U$-index ranked countries which are ranked highly have large population sizes. The only ranking scheme shown that places higher priority on larger NOCs is the total medals based ranking in gray. The per-capita based ranking generally gives priority to smaller NOCs. The Bayesian ranking can be reasonably described as achieving a balance  between smaller and larger nations, with a mix of such nations achieving high ranks.

\subsection{Long-term Stability of Bayesian Ranking}
\label{sec:results-stability}

One of the key motivators for the Bayesian ranking approach is to provide a stable long-term estimate of each NOC's performance, based on a single Olympic Games, using shrinkage effects whilst also accounting for population size effects. Section~\ref{sec:results-shrinkage} demonstrated the shrinkage effects and how they approach accounts for population size. Here we will evaluate its ability to provide an interpretable long-term measure of performance by considering its stability and behavior when applied to many past Olympic Games.

\begin{figure}[!ht]\centering
\includegraphics[width=\linewidth]{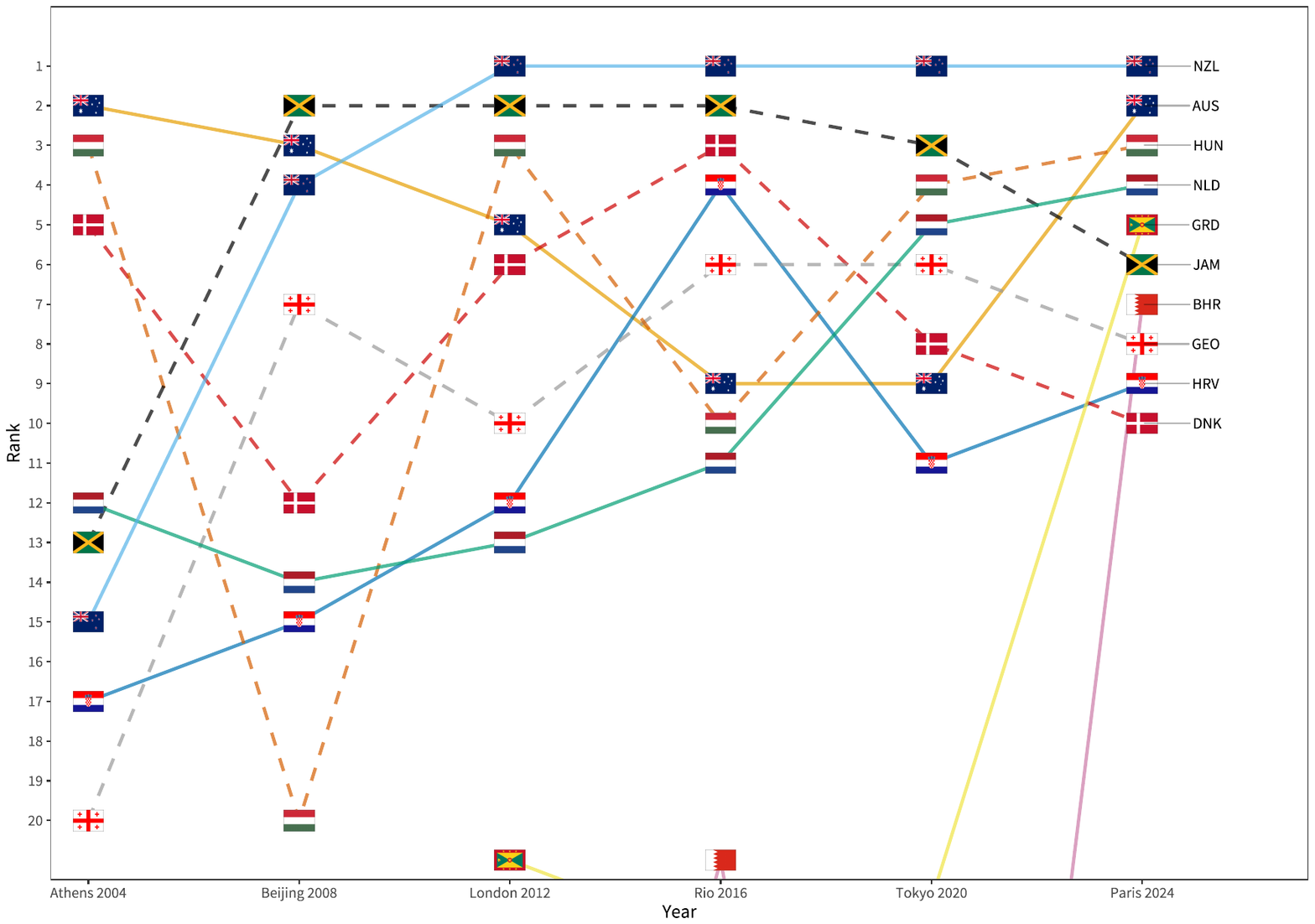}
\caption{Top 10 ranked countries across the last 6 Olympic Games. Every country that is within the Top 10 in any Game is joined to their rank across all 6 games.}
\label{fig:history}
\end{figure}

The Bayesian ranking method is applied entirely separately for each of the last 6 Olympic Games, so there is no carryover of information across Olympic Games in the inference. Notice the strong consistency in the Bayesian ranking for each NOC shown across these 6 Games, which provides comfort in the interpretation of the ranks as a stable long-term estimate of each NOC's performance, due to the shrinkage effects. In particular, New Zealand topped the ranks in the last 4 Olympic Games after a notable improvement since Athens 2004 which is known to be influenced by increases in funding and strategy/policy priority changes resulting from the creation of Sparc (now Sports NZ) and its subsidiary High Performance Sports NZ which were formed in 2002 and 2012 respectively. The Netherlands have made consistent gains since 2008. Hungary did not perform so well in Beijing 2008, but otherwise ranked very highly. Similarly, Georgia has performed consistently well. Jamaica dropped from a consistent rank of 2-3 in size Beijing 2008 to rank 6 in Paris 2024. Interestingly, both Grenada and Bahrain performed sufficiently well in Paris 2024 to appear in the Top 10 for the first time, despite the shrinkage applied to their estimates.

\section{Discussion}
\label{sec:discussion}

In this final section, we discuss some of the assumptions behind our model, possible ways it might be extended and more detailed comparison to the \cite{jrnl:duncan2024population} method. We first examine the equal treatment of medal types and the validity of the Poisson assumption. We then discuss the assumption of constant conditional probabilities for multi-medal winners across countries. Finally, we compare our Bayesian ranking algorithm to existing probabilistic approaches, particularly the Duncan-Parece method, and highlight key advantages of our approach in capturing long-term national Olympic performance.

Throughout the analysis we have treated all medals equally, but we could have considered either separate Poisson processes for each type of medal or having a weighting scheme for each type of medal. The former medal type specific Poisson process would have many zeroes and therefore limit the information to the likelihood. In this case, the inference would be strongly influenced by the priors which is best avoided. Previous literature has justified treating gold, silver and bronze medals equally. For instance, \cite{jrnl:bian2005predicting} has argued that the performances winning gold, silver and bronze medals are often very similar, with a degree of luck being the crucial determining factor, which partially justifies equal weighting. \cite{jrnl:ogwang2021olympic} also found that the first principal component estimated from the joint distribution of gold, silver and bronze medal totals across countries had approximately equal weight on the number of gold, silver and bronze medals and explains over 94\% of the total variance, indicating very strong correlation in the gold, silver and bronze medal counts across different countries. 

The Poisson assumption, as a limiting approximation to the binomial distribution for the number of unique medal winners per country, can be empirically tested using observed medal counts over multiple Olympic Games. Assuming that a country's true medal-winning rate remains stable across successive Olympic cycles, we have considered diagnostic plots of estimated mean and variance, and hypothesis tests to detect possible excess-Poisson variation (see the online Appendix). These tests provide support for the Poisson approximation. While there are obvious physical limitations to the binomial argument that total medals can be regarded as a sum of equal probability Bernoulli trials, these are likely negligible in practice. For example, factors such age, socioeconomic conditions, and race may influence the likelihood of individual success in different events. However, even if individual probabilities vary the total medal counts can still be reasonably modeled using a Poisson distribution (with rate scaled by population size) as long as individual successes or failures can be treated as independent events.  The assumption of independence is somewhat more difficult to justify. On one hand, given the rarity of athletes with the talent to win an Olympic medal, it seems reasonable to treat these individuals as independent rare events. On the other hand, the notion that `a rising tide lifts all boats' should be acknowledged - training environments, national sports culture, and psychological factors can create performance spillover effects, particularly in elite competitive sports. 

Recall the model assumes that the conditional probabilities of an individual winning two or more medals, given that they have won at least one medal are constant across all countries. We have partially justified this assumption under the premise that when conditioning on an Olympic medal winner, one is implicitly conditioning on having the training environment, funding and genetic talent to compete as a world class athletes (i.e. we are mostly conditioning away those factors that might cause differing levels of success per-capita across countries). However, if a country focuses their resources on particular sports (such as swimming) where multi-medal winners are more likely, the true conditional probabilities for that country of winning multiple medals given a single medal may be larger than the global average.  Unfortunately, there is not enough data from a single Olympic Games to estimate these conditional probabilities in a coherent way at a per-country level, particularly for smaller countries and, as such, complete pooling represents an attractive modeling assumption from a practical standpoint.  In practice, the effects of incorporating such corrections in our ranking (which would in any case be driven primarily by single medal winners who constitute the vast majority of medals) would be unlikely to change the conclusions. To elaborate further, since the conditional probabilities are not country specific, the shrinkage applied to a country's raw per-capita medal rate by our model will tend to be slightly larger if its medals were won by disproportionally few athletes compared to a scenario where the same number of medals were won by separate athletes.  This effect may have resulted in the USA doing slightly worse in the Bayesian ranking for Paris 2024 (position 51) compared to the observed per-capita ranking (position 48).  However, as shown in Figure~\ref{fig:credrank}, the practical differences in performance between countries ranked as 51 and 48 by Bayesian ranking is statistically negligible due to the strongly overlapping credible intervals.

To produce a ranking, we have summarized the posterior distribution of the rank for a country's medal rate by its mean.  As demonstrated in Figure~\ref{fig:credrank}, the posterior distributions of highly-placed smaller countries tend to be quite right skewed compared to larger countries, with the result that they may be ranked less advantageously using the posterior mean rank compared to using the posterior median rank.  For instance, Grenada would be ranked as the 2nd best country in Paris 2024 according to posterior median rank, as opposed to 5th with posterior mean rank, although the other positions in the top 10 are identical to those in Table~\ref{tab:ranks}, having accounted for that shift.  Effectively, the posterior mean rank applies a relative penalty to the ranking of smaller countries with asymmetric high-variability posterior distributions, compared to countries that have symmetric posteriors.  We have chosen the posterior mean rank as it is in keeping with previous applications of Bayesian ranking, such as \cite{jrnl:laird1989empirical}, but posterior median rank could equally be used to create the ranking table.  Alternative rankings (such as rankings based on the posterior median) can be explored using the Shiny app at \cite{web:shiny}

The goal of our model is to infer which countries are most efficient in producing Olympic medalists relative to their population size.  This population-scaled efficiency is a subjective definition of success, and may not be something that a country strives for given that some Olympic events have more prestige and glamour than others.  For instance, while the United States places at 51 on our Bayesian ranking table (despite winning the table on standard lexicographic medal counts), it does particularly well in many of the `marquee' track and field events that enjoy particular media attention.  Many factors might influence this relative Olympic success such as GDP per capita, the proportion of GDP allocated to sports funding, climate, demographic and social factors.  If the aim was to instead infer which countries were most successful given fixed values of such variables, such factors could be accounted for in a Bayesian Poisson regression, which again used log-population size as an offset in modeling the probability of a unique medal, and incorporated a normal random effect for each country, with the inferred value of this random effect used to rank differing NOCs.  The implicit prior for medal winning probabilities across countries would be log-normal (as opposed to beta) in this modified model, but with shifts in the mean parameter based on the covariate information for a NOC.

This brings up another point of discussion.  How much influence does the hierarchically estimated beta prior influence the overall results and ranking?  The choice of beta rather than log-normal or some other choice is a subjective choice, and could be considered as a weakness of our model by some.  It turns out that the exact choice of hierarchical prior has relatively little influence in the final results.  We have also investigated hierarchical specifications of log-normal, logit normal and mixture-beta priors for the probability of a unique medal (see the online Appendix for details).  In all models, New Zealand is ranked as the best performing NOC, and the top 5 and top 10 ranked countries are consistent across all models, with slightly differing orderings based on the implied shrinkage.  In particular, the results from the mixture beta prior are almost identical to the results above.  Given that the mixture beta has theoretical advantages in that it can adapt to the most general distribution of probabilities across countries, it is comforting that the results from the beta and mixture beta priors are so similar.

Finally we return to the approach of \cite{jrnl:duncan2024population}, which ranks by the likelihood of achieving a country's observed medal count (or higher) under a binomial model.  That is, the model considers possible awarded medals to a country as independent Bernoulli trials with probability success proportional to the country's population.   One nice feature of their approach is their binomial probability implicitly respects the total quota of possible Olympic medals that a country can win.  For instance, in Tokyo 2021 the total medals that a country could possibly win was 579 due to restrictions on the total number of athletes that an NOC could enter.  While this isn't respected by our Poisson distribution (which has infinite support), we note that in practice the probability distributions for total medals corresponding to Poisson and binomial approaches are very similar again based on the large $n$ and small $p$ approximation (See the online Appendix for more details).

The Duncan-Parece method is effectively a transformed $p$-value that tests the null hypothesis that the probability of a country winning a particular Olympic medal equals the proportion of the aggregate population of countries that have previously won an Olympic medal that resides in that country. The index tends to favorably rank higher population countries.  For example, in the Tokyo 2020 Olympics, the USA ranked 6th using Duncan-Parece’s $U$-index, with a score of 15.2.  However, if we imagine that both the population of the USA and its medal count doubled, the $U$-index for the USA increases to 34, in which case they would would be ranked as the top performing country.  This is despite the USA’s relative performance (in terms of medals per-capita) remaining unchanged.  The effect just described is due to a well known feature of $p$-values \citep{jrnl:wasserstein2016asa}.  Small observed effects (such as small observed differences between country-specific and global per-capita medal ratio) can correspond to extremely small $p$-values when the effect is estimated with small standard error (roughly the standard error for estimated per-capita ratios are proportional to the square root of population).  These issues have been previously discussed in the context of ranking effect sizes in genetics and genomics in \cite{jrnl:ferguson2020empirical}, and sheds some light on the differences in rankings produced by our Bayesian ranking algorithm and their probabilistic index.  To advocate for our approach, as demonstrated in Section~\ref{sec:results}, New Zealand has led our Bayesian ranking algorithm consistently over the previous four Olympic cycles, and they have also `topped' the empirical per-capita medal table over this collective period when restricting to countries that have a population of at least 1,000,000; indicating that our algorithm is better at inferring the countries with the best long-run performance even though we infer this performance from a single Olympic Games.

\section{Conclusion}
\label{sec:conclusion}

In conclusion, our Bayesian ranking algorithm is a new approach way to rank each country's performance in international competitions such as the Olympic Games. It strikes a balance between consideration of total medals, which would give an advantage to larger countries, and empirical per-capita medal ratios which are overly influenced by stochastic factors pertaining to a particular Olympics and notable advantage the large number of smaller nations.

\bibliographystyle{chicago}
\bibliography{Bayesian_Ranking}

\clearpage
\appendix
\section*{Appendix}
\addcontentsline{toc}{section}{Appendix}

In this appendix we consider secondary details about the method and results. A Shiny app is available from \cite{web:shiny} for interactively exploring the results of the Bayesian ranking, including how the ranking might vary by country-specific demographic factors and comparison to other ranking methods.

\section{Bayesian Shrinkage and Ranking Stability}
\label{sec:shrinkage}

In the main paper we presented the shrinkage results for the medal winning countries in Paris 2024. The shrinkage brings each country closer to the global average, as a means to provide a more stable long-run estimate of their performance. The large countries see little shrinkage as they are well informed by the data, while the smaller nations see more shrinkage. The countries with one or more multi-medal winning athletes also saw more shrinkage than those with only single medal winning athletes of a similar size, as the multi-medal winning rate is assumed constant across all nations. Here we will explore the (reverse) impact of the shrinkage of the countries who won no medals in Paris 2024

In the main paper we presented the shrinkage results for the medal winning countries in Paris 2024. The shrinkage brings each country closer to the global average, in order to provide a more stable long-run estimate of their performance. The large countries see little shrinkage as they are well informed by the data, whereas the smaller nations see more shrinkage. The countries with one or more multi-medal winning athletes also saw more shrinkage than those with single medal winning athletes of a similar size, as the multi-medal winning rate is assumed constant across all nations. Here we will explore the (reverse) impact of the shrinkage of the countries who won no medals in Paris 2024.

\begin{figure}[!ht]\centering
\includegraphics[width=\linewidth]{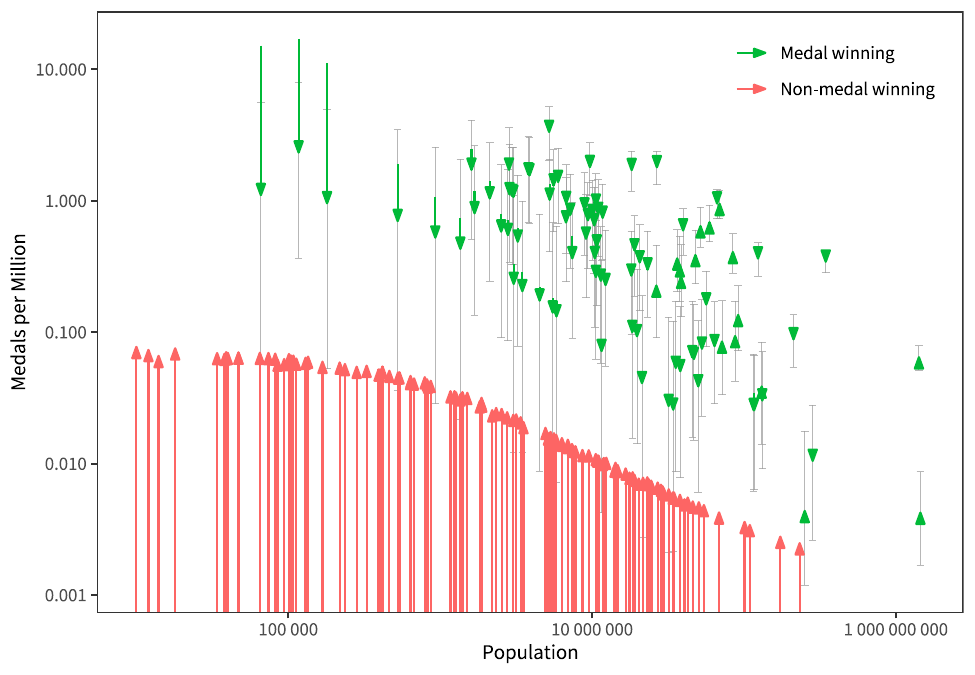}
\caption{Effect of shrinkage on the medal winning rate against population size for each country that won a medal (as per Figure 1 of the main paper, and the inflation effect for those who won no medals.}
\label{fig:upshrink}
\end{figure}

The uplift of the medal winning rate for NOCs with no medalists can be seen in Figure~\ref{fig:upshrink} by the vertical arrows. The shrinkage for these NOCs is only affected by their population size, as they have zero medal count and no multi-medal winners. Broadly speaking, there is a monotonically decreasing inflation effect of the shrinkage with population size. The smallest countries receive the biggest uplift towards the global average, as the larger populations' estimates are more data-driven.

The uplift of the medal winning rate for NOCs with no medalists can be seen in Figure~\ref{fig:upshrink} by the vertical arrows. The shrinkage for these NOCs is only affected by their population size, as they have zero medal count and no multi-medal winners. Broadly speaking, there is a monotonically decreasing inflation effect of the shrinkage with population size. The smallest countries receive the biggest uplift towards the global average, as the larger populations' estimates are more data-driven. The non-medal winning countries are not assigned a rank in the Bayesian ranking, as per usual protocol.

\begin{eqnarray*}
  E(M_c/n_c)=p_c\,(1-q_2 + 2\,q_2\,(1-q_3) + 3\,q_2\,q_3\,(1-q_4) + 4\,q_2\,q_3\,q_4)\approx p_c
\end{eqnarray*}

The non-medal winning countries are not assigned a rank in the Bayesian ranking, as per usual protocol. However, if they were ranked they would be amongst the bottom countries, which can be seen by the low height of their arrows compared to most medal winning NOCs. 

\section{Approximate Equivalence to Duncan-Parece Method}
\label{sec:equivalence-dp}

The model underlying the proposed Bayesian ranking can be considered as an approximation to the \cite{jrnl:duncan2024population} method in the circumstance that the probabilities $p_c$, which we define in the Methods section of the main manuscript, are equal across competing NOCs. The Duncan-Parece method models the number of medals won by a NOC using a binomial distribution where each trial is one of the medals that an NOC could win, up to the maximum possible number of medals, with a country specific probability of winning. Under the quota system the maximum number of medals that could be achieved was $M=559$ trials for Paris 2024, out of a total of $T=1039$ medals. Under the assumption of equal medal winning capability for any individual around the world, the expected number of medals won by a country is proportional to the country's population (see equation (4) of \cite{jrnl:duncan2024population}). But due to the close equivalence between the Poisson/binomial, you can view our per-capita specification as simply changing the denominator term.

The model underlying the proposed Bayesian ranking can be considered as an approximation to the \cite{jrnl:duncan2024population} method in the circumstance that the probabilities $p_c$, which we define in the Methods section of the main manuscript are equal across competing NOCs. The Duncan-Parece method models the number of medals won by a NOC using a binomial distribution where each trial is one of the medals that an NOC could win, up to the maximum possible number of medals, with a country specific probability of winning. Under the quota system the maximum number of medals that could be achieved was $M=559$ trials for Paris 2024, out of a total of $T=1039$ medals.   Under the assumption of equal medal winning capability for any individual around the world, the expected number of medals won by a country is proportional to the country's population (see equation (4) of \cite{jrnl:duncan2024population}). But due to the close equivalence between the Poisson/binomial, you can view our per-capita specification as simply changing the denominator term of the binomial probability.

The population size for country $c$ is $n_c$ and the total population of the world is $N$. The Duncan-Parece model is that the number of medals won by any country is $M_c\sim\text{Binomial}(M, \pi_c)$ where $\pi_c$ is probability of that country winning a given medal.  Under the equal capability assumption (per person) across all NOCs, the expected number of medals won should be proportional to that countries population so $M\times \pi_c=\frac{n_c}{N}T$. Rearranging we obtain:
$$\pi_c=\frac{n_c}{N}\frac{T}{M}$$
so under the equi-capability reference model the number of potential medals follows a Binomial$\left(M, \frac{n_c}{N}\frac{T}{M}\right)$. Such a binomial is approximately a Poisson$\left(\frac{n_c}{N}T\right)$ which in turn is approximately Binomial$\left(n_c, \frac{T}{N}\right)$. The latter probability $p_c=T/N$ is  per-capita probability of an individual from a given NOC winning a medal underlying our model, assuming it does not change over countries (that is $p_c$ is constant).

Although the two underlying models approximate each other under the null equi-capability reference model, our inference approach does not have the same restriction on small countries due to the sample size impacts on the (transformed) $p$-values which underlie the $U$-index of \cite{jrnl:duncan2024population}. Another substantial improvement in our method is the explicit handling of multi-medal winning athletes, which would break the usual independence assumption of the binomial. A drawback of our approach compared to \cite{jrnl:duncan2024population} is that we do not explicitly handle each country's quota. However, even for the most successful countries, medal winning is a rare event so no country get sufficiently close to their quota for this to be of concern.

\section{Sensitivity to Prior Choice}

\begin{table}
\begingroup\fontsize{8}{10}\selectfont
\resizebox{\linewidth}{!}{%
\begin{tabu} to \linewidth {
  >{\raggedright}X[0.4]   
  >{\raggedright}X[0.4]   
  >{\raggedleft\arraybackslash}X[1.6]    
  >{\raggedleft}X[0.7]    
  >{\raggedleft}X[0.7]    
  >{\raggedleft}X[0.7]    
  >{\raggedleft\arraybackslash}X[0.9]   
  >{\raggedleft\arraybackslash}X[1.1]   
  >{\raggedleft\arraybackslash}X[1.6]   
  >{\raggedleft\arraybackslash}X[1.6]   
  >{\raggedleft\arraybackslash}X[0.8]   
  >{\raggedleft\arraybackslash}X[1]   
}

\toprule
\multicolumn{3}{c}{\textbf{Country}} & 
\multicolumn{3}{c}{\textbf{Posterior Mean Ranks}} & 
\multicolumn{4}{c}{\textbf{Medals per million}} & 
\multicolumn{2}{c}{\textbf{Beta CI}} \\
\cmidrule(lr){1-3} \cmidrule(lr){4-6} \cmidrule(lr){7-10} \cmidrule(lr){11-12}

 &  & Population (thousands)& Beta & Log-Normal & Mixed-Beta & Observed & Beta Median & Log-Normal Median & Mixed-Beta Median & 95\% Lower & 95\% Upper\\

\midrule
\includegraphics[height=1em, width = 1.5em]{nz.png} & NZL & 5,214 & 1 & 1 & 1 & 3.84 & 3.31 & 3.61 & 3.27 & 2.05 & 4.89 \\
\includegraphics[height=1em, width = 1.5em]{au.png} & AUS & 26,713 & 2 & 3 & 2 & 1.98 & 1.80 & 1.82 & 1.79 & 1.32 & 2.31 \\
\includegraphics[height=1em, width = 1.5em]{hu.png} & HUN & 9,676 & 3 & 4 & 3 & 1.96 & 1.83 & 1.84 & 1.79 & 1.06 & 2.89 \\
\includegraphics[height=1em, width = 1.5em]{nl.png} & NLD & 18,229 & 4 & 5 & 4 & 1.87 & 1.68 & 1.73 & 1.71 & 1.15 & 2.40 \\
\includegraphics[height=1em, width = 1.5em]{gd.png} & GRD & 117 & 5 & 2 & 6 & 17.09 & 2.42 & 7.14 & 2.13 & 0.34 & 8.16 \\
\includegraphics[height=1em, width = 1.5em]{jm.png} & JAM & 2,839 & 6 & 7 & 5 & 2.11 & 1.71 & 1.82 & 1.68 & 0.67 & 3.54 \\
\includegraphics[height=1em, width = 1.5em]{bh.png} & BHR & 1,607 & 7 & 6 & 7 & 2.49 & 1.74 & 2.00 & 1.66 & 0.53 & 4.25 \\
\includegraphics[height=1em, width = 1.5em]{ge.png} & GEO & 3,808 & 8 & 8 & 8 & 1.84 & 1.58 & 1.68 & 1.56 & 0.65 & 3.00 \\
\includegraphics[height=1em, width = 1.5em]{hr.png} & HRV & 3,875 & 9 & 9 & 9 & 1.81 & 1.60 & 1.65 & 1.54 & 0.65 & 3.10 \\
\includegraphics[height=1em, width = 1.5em]{dk.png} & DNK & 5,977 & 10 & 10 & 10 & 1.51 & 1.35 & 1.42 & 1.40 & 0.64 & 2.50 \\
\includegraphics[height=1em, width = 1.5em]{ie.png} & IRL & 5,255 & 15 & 15 & 15 & 1.33 & 1.01 & 1.05 & 1.02 & 0.42 & 2.04 \\
\includegraphics[height=1em, width = 1.5em]{fr.png} & FRA & 66,549 & 16 & 16 & 16 & 0.96 & 0.95 & 0.95 & 0.95 & 0.72 & 1.22 \\
\includegraphics[height=1em, width = 1.5em]{gb.png} & GBR & 69,138 & 17 & 19 & 17 & 0.94 & 0.95 & 0.95 & 0.95 & 0.72 & 1.21 \\
\includegraphics[height=1em, width = 1.5em]{us.png} & USA & 345,427 & 51 & 47 & 51 & 0.36 & 0.34 & 0.34 & 0.34 & 0.29 & 0.41 \\
\includegraphics[height=1em, width = 1.5em]{cn.png} & CHN & 1,419,321 & 76 & 78 & 77 & 0.06 & 0.06 & 0.06 & 0.06 & 0.05 & 0.08 \\

\bottomrule
\end{tabu}
}
\endgroup

\caption{
Bayesian ranking of the top 10 and selected countries in Paris 2024 based on posterior estimates of medal-winning probabilities, normalized per million population. Rankings are shown under different prior specifications: the beta prior, a log-normal prior, and a flexible mixed-beta prior. The ranks are based on mean ranking of posterior mean probabilites. Corresponding 95\% credible intervals for the beta prior are also provided supporting the appropriateness of the beta prior in the analysis.
}

\label{tab:prior_sensitivity}
\end{table}

\begin{figure}[!ht]
\centering
\includegraphics[width=\linewidth]{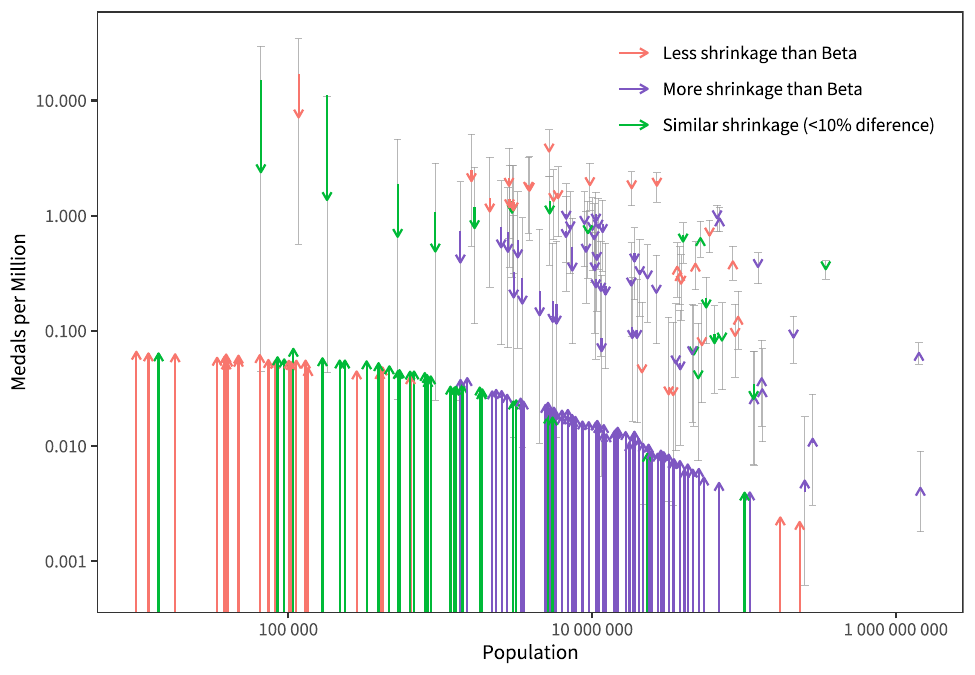}
\caption{Effect of shrinkage on the medal-winning rate against population size for each country that won a medal (as per Figure 1 of the main paper), and the inflation effect for those who won no medals. for a log-normal prior with colours indicating how the shrinkage compares under each model}
\label{fig:lognorm_prior}
\end{figure}

The shrinkage which stabilizes the long-run estimates of expected medal winning performance is mostly imposed by the choice of prior distribution (with only mild impacts from the assumed common multi-medal winning probabilities across all NOCs). In the main paper, we present the results when a beta prior distribution is used to inform the country-specific probability of an individual becoming a unique medal winner, $p_c=P(X_c\geq 1),$:
$$p_c\sim \textrm{Beta}(\alpha, \beta)$$
with hyperparameters:
\begin{eqnarray*}
    \alpha&\sim& \textrm{Uniform}(0, 1), \textrm{and}\\
    \beta&\sim& \textrm{Uniform}(0, 10^8).
\end{eqnarray*}
This prior is a natural choice and appears to produce sensible Bayesian rankings, but we have also explored the sensitivity to alternative priors and their hyperparameters.

Firstly, we considered a truncated log-normal prior distribution which allows for a heavier upper tail for the prior distribution of the unique medal winner probability
$$p_c\sim \textrm{Log-Normal}(\mu, \sigma) I(0,1)$$
with hyperparameters:
\begin{eqnarray*}
    \mu&\sim& \textrm{Normal}(-15, 0.323), \textrm{and}\\
    \sigma&\sim& \textrm{Gamma}(0.001, 0.001)
\end{eqnarray*}
where the gamma is parameterized by the shape and rate. The hyperprior for the mean is chosen to be close to the overall medal winning rate over the world population. The resulting shrinkage is shown in Figure~\ref{fig:lognorm_prior}. The shrinkage is somewhat flatter compared to the beta prior, so that the smaller countries like Grenada are not shrunk as much as for the beta prior, and the larger countries experience notably more shrinkage. The trajectory of shrinkage vs population for the non-medal winning countries is also notably flatter compared to with the beta prior.  
The resultant rankings are shown in Table~\ref{tab:prior_sensitivity}. The beta prior is used as the reference, where a black font means no change, green is an improvement in the ranked position and orange is a decline. The resulting ranks do not change much compared to the beta prior. The key difference in the Top 10 is that Grenada moves from rank 5 with the beta prior to 2nd place. But notice that the shrinkage for the larger countries leads to the UK being slightly pulled down the ranks, and the USA and Japan brought slightly up the ranks.

The shrinkage for larger countries is indicative of the prior being informative over the large amount of sample information, which is not desirable. In addition, it appears that the heavy tail of the log-normal is allowing substantial variation in the probability of being a unique medal winner, and subsequently higher posterior mass on extreme performance, for countries with small populations. However, when one considers the uncertainty via the 95\% credible intervals for the medal winning rate (calculated from the beta specification) these changes are not significant, as the estimated rates from the log-normal prior are well within these intervals (although it is worth noting that the medal winning rate for Grenada is close to the upper bound).  

A logit-normal prior:
$$p_c\sim \textrm{Logit-Normal}(\mu, \sigma)$$
with hyperparameters:
\begin{eqnarray*}
    \mu&\sim& \textrm{Normal}(-15, 0.323), \textrm{and}\\
    \sigma&\sim& \textrm{Gamma}(0.001, 0.001)
\end{eqnarray*}
gave near identical results to the log-normal, so are not discussed for brevity.

\begin{figure}[!ht]
\centering
\includegraphics[width=\linewidth]{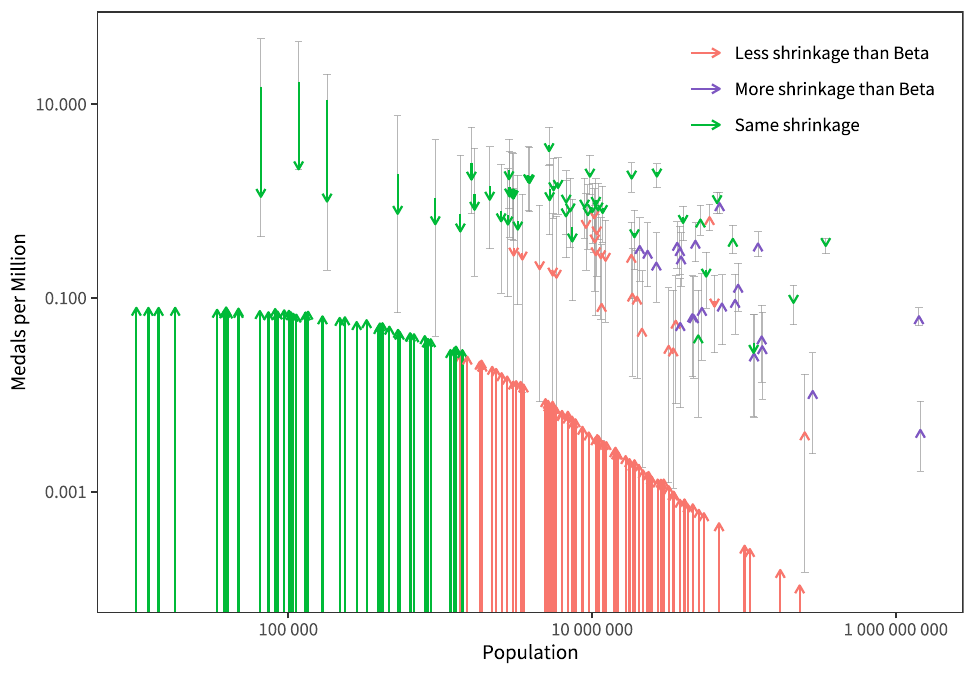}
\caption{Effect of shrinkage on the medal-winning rate against population size for each country that won a medal (as per Figure 1 of the main paper), and the inflation effect for those who won no medals. for a mixed-beta prior with colours indicating how the shrinkage compares under each model}
\label{fig:mixbeta_prior}
\end{figure}

A very flexible mixture of beta components was also explored to assess the robustness of the prior to heavy-tailed behavior. Three components were used, offering enough flexibility to approximate characteristics of priors such as the log-normal, while including the standard beta prior as a special case. The three components
$$p^j_c\sim \textrm{Beta}(\alpha_j, \beta_j)$$
for $i=1, 2, 3$ have the same hyperparameters as for usual single beta prior as above. A Dirichlet(1,1,1) distribution is used for the mixture weights for the three beta components. The resulting shrinkage is shown in Figure~\ref{fig:lognorm_prior} is very similar to using simple one component beta prior which provides comfort in the results presented the main paper, as it appears the choice of beta prior (and hyperparameters) is not strongly influential on the results and the changes seen with the log-normal prior above are likely due to this prior specification enforcing a unrealistically heavy tail that fails to adequately shrink highly stochastic extreme rates for small NOCs.

One subtle distinction, however, emerges at the tails: at large population sizes for non-medal winners, the mixed-beta prior results in slightly less shrinkage than the regular beta. This can occur because the mixture model may allocate higher-population observations more weight from flatter components, which exert less shrinkage. However, these deviations are minimal and affect only non-medal-winning countries, which are excluded from the final rankings.

\section{Poisson assumptions}

\begin{figure}[!ht]\centering
\includegraphics[width=\linewidth]{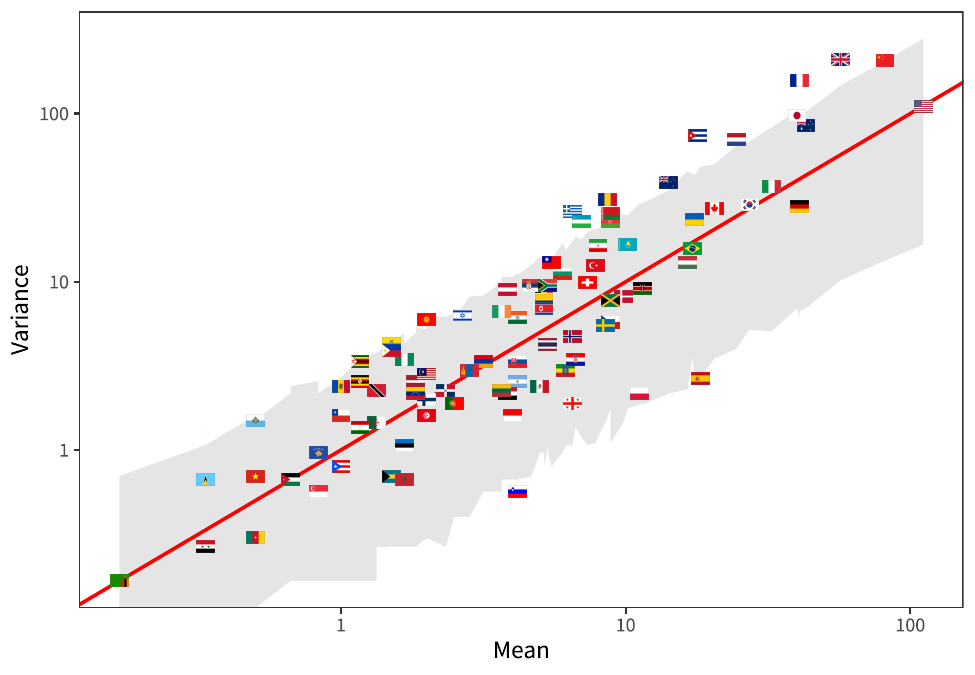}
\caption{Average medals over the seven previous Olympic games are plotted against the empirical variance of the number of medals (y-axis), by country.  Gross departures from the line of equality signify possible violations of the Poisson assumption. The shaded area represents a 90\% predictive band for the empirical variance of a random sample of size seven from the Poisson distribution.}
\label{fig:meanvar}
\end{figure}

Using the medal winning results across multiple Olympic Games we can evaluate the equal mean/variance assumption of the Poisson model. Figure~\ref{fig:meanvar} plots the sample mean against the variance of the number of medals for each NOC across all seven of the Olympic Games from Athens 2004 to Paris 2024, shown on a log-scale. A non-medal winning country is included as a zero for that Games, but a country which did not compete in the Games are set to missing. You can clearly see a strong mean-variance relationship as the NOCs are close to the line of equality. Some of the notable deviations from the line of equality are due to host NOCs (Greece, China, UK, Brazil, Japan and France). But overall the equal mean/variance assumption of the Poisson appears extremely reasonable.

There is a mild crudeness to this evaluation due to the multi-medal winning athletes. In our model, the total number of medals won by country $c$ is $M_c=1 \times M_{1,c}+2\times M_{2,c}+\cdots\sum_i i\,M_{i.c}$. Each of the $M_{i,c}$ is assumed to be Poisson distributed. But the total number of medals won $M_c$ is not exactly Poisson due to the scaling by the number of medals $i$, but will be close to Poisson as the multi-medal winning athletes are sufficiently rare especially as the total $M_c$ will be dominated by the $1\times M_{1,c}$ term which is exactly Poisson. 

Another aspect of approximation in this diagnostic plot is that it assumes the number of medals on offer (including quotas of those available), population size, and the underlying performance of each NOC (as measured by their long-term medal rate: $E(M_c/n_c)$) are constant over time. As we are considering only Games following Athens 2004 these factors will not have changed much, and so is of little concern.

As an alternative test of the Poisson assumption, we defined p-values to test for excess Poisson variation in the medal counts over two successive Olympic Games, assuming performance, the total number of medals on offer and population all remain stable.  Under these assumptions (in addition to independence) conditional on $R$, that is the total medals won over two sucessive games, the number of medals won by a country in a particular Games has a binomial distribution with $R$ trials and probability of success 0.5.  The results are not shown here, but the country-specific p-values demonstrated no evidence for super-Poisson variability (with the notable exception of host countries).

\end{document}